\pdfoutput=1
\documentclass[aps,prd,amsmath,floats,floatfix, twocolumn,
superscriptaddress,nofootinbib,showpacs,longbibliography]{revtex4-1}
\usepackage[T1]{fontenc}
\usepackage[utf8]{inputenc}
\usepackage{lmodern}
\usepackage{verbatim}
\usepackage[dvipsnames, usenames]{xcolor}
\definecolor{linkcolor}{rgb}{0.0,0.3,0.5}
\usepackage[hypertexnames=false, unicode, colorlinks=true, linkcolor=linkcolor,
citecolor=linkcolor, filecolor=linkcolor,urlcolor=linkcolor,
pdfusetitle]{hyperref}
\usepackage[all]{hypcap}
\usepackage{graphicx}
\usepackage{xspace}
\usepackage{amssymb}
\usepackage{microtype}
\usepackage[english]{babel}
\usepackage{blindtext}
\usepackage[normalem]{ulem} %for \sout
\usepackage{bm} % boldmath
\usepackage{mathrsfs}

\graphicspath{{figs/}}

\usepackage[nomessages]{fp}

\begin{document}

\title{Interplay between numerical-relativity and black hole perturbation theory \\in the intermediate-mass-ratio regime}
\newcommand{\UMassDMath}{\affiliation{Department of Mathematics,
		University of Massachusetts, Dartmouth, MA 02747, USA}}
\newcommand{\UMassDPhy}{\affiliation{Department of Physics,
		University of Massachusetts, Dartmouth, MA 02747, USA}}
\newcommand{\CSCVRUMass}{\affiliation{Center for Scientific Computing and Visualization Research, University of Massachusetts, Dartmouth, MA 02747, USA}}

\author{Tousif Islam}
\email{tislam@umassd.edu}
\UMassDPhy
\UMassDMath
\CSCVRUMass

% Because hyperref only gets the *last* author, we need to be explicit.
\hypersetup{pdfauthor={Islam et al.}}
\date{\today}
%==========================================================================
\begin{abstract}
We investigate the interplay between numerical relativity (NR) and point-particle black hole perturbation theory (ppBHPT) for quasi-circular non-spinning binary black holes in the intermediate mass ratio regime: $7 \le q \le 128$ (where $q:=m_1/m_2$ is the mass ratio of the binary with $m_1$ and $m_2$ being the mass of the primary and secondary black hole respectively).
Initially, we conduct a comprehensive comparison between the dominant $(\ell,m) = (2,2)$ mode of the gravitational radiation obtained from state-of-the-art NR simulations and ppBHPT waveforms along with waveforms generated from recently developed NR-informed ppBHPT surrogate model, \texttt{BHPTNRSur1dq1e4}. This surrogate model employs a simple but non-trivial rescaling technique known as the $\alpha$-$\beta$ scaling to effectively match ppBHPT waveforms to NR in the comparable mass ratio regime. 
Subsequently, we analyze the amplitude and frequency differences between NR and ppBHPT waveforms to investigate the non-linearities, beyond adiabatic evolution, that are present during the merger stage of the binary evolution and propose fitting functions to describe these differences in terms of both the mass ratio and the symmetric mass ratio. Finally, we assess the performance of the $\alpha$-$\beta$ scaling technique in the intermediate mass ratio regime.
\end{abstract}

\maketitle
%==========================================================================
%==========================================================================
%==========================================================================

\section{Introduction}
The detection and characterization of gravitational wave (GW) signals from binary black hole (BBH) mergers require computationally efficient yet accurate multi-modal waveform models. The development of such models relies heavily on accurate numerical simulations of BBH mergers. The most accurate way to simulate a BBH merger is by solving the Einstein equations using numerical relativity (NR). Over the past two decades, NR pipelines have been refined for BBH systems with comparable masses ($1 \le q \le 10$)~\cite{Mroue:2013xna,Boyle:2019kee,Healy:2017psd,Healy:2019jyf,Healy:2020vre,Healy:2022wdn,Jani:2016wkt,Hamilton:2023qkv}. The availability of a substantial number of NR simulations in the comparable mass ratio regime has facilitated the development of computationally efficient and accurate approximate models, such as reduced-order surrogate models based on NR data \cite{Blackman:2015pia,Blackman:2017pcm,Blackman:2017dfb,Varma:2018mmi,Varma:2019csw,Islam:2021mha}, or semi-analytical models calibrated against NR simulations \cite{bohe2017improved,cotesta2018enriching,cotesta2020frequency,pan2014inspiral,babak2017validating,husa2016frequency,khan2016frequency,london2018first,khan2019phenomenological}.
On the other hand, extreme mass ratio binaries (i.e. $q \rightarrow \infty$) can, in principle, be modelled accurately with point particle black hole perturbation theory (ppBHPT) where the smaller black hole is treated as a point particle orbiting the larger black hole in a curved space-time background. Substantial progress has been made over the past two decades in simulating BBH mergers accurately in this regime~\cite{Sundararajan:2007jg,Sundararajan:2008zm,Sundararajan:2010sr,Zenginoglu:2011zz,Fujita:2004rb,Fujita:2005kng,Mano:1996vt,throwe2010high,OSullivan:2014ywd,Drasco:2005kz}. 

However, it is the intermediate mass ratio regime ($10 \leq q \leq 100$) that still presents significant challenges for performing accurate simulations of BBH mergers. NR simulations for binaries in this mass ratio range become exceedingly computationally expensive for a variety of reasons. On the other hand, as the binary becomes less asymmetric, the assumptions of the ppBHPT framework begin to break down. Therefore, the intermediate mass ratio regime provides a unique opportunity to compare and contrast results obtained from NR and ppBHPT framework. In particular, Refs.~\cite{Lousto:2010tb,Lousto:2010qx,Nakano:2011pb} studied this regime to gain insights into the limitations and accuracy of both approaches as well as to further the understanding about the dynamics of the binary. 

Recently, a significant milestone has been reached with the development of the \texttt{BHPTNRSur1dq1e4} surrogate model~\cite{Islam:2022laz}. This model, based on the ppBHPT framework, accurately predicts waveforms for comparable to large mass ratio binaries. Through a simple but non-trivial calibration process, the ppBHPT waveforms are rescaled to achieve a remarkable agreement with NR data in the comparable mass ratio regime. 
In a parallel effort, Ref. \cite{Wardell:2021fyy} has developed a fully relativistic second-order self-force model, which also demonstrates excellent agreement with NR in the comparable mass ratio regime.
Additionally, recent advancements in NR techniques have pushed the boundaries of BBH simulations, enabling the simulations of BBH mergers with mass ratios up to $q=128$ for various spin configurations \cite{Lousto:2020tnb,Lousto:2022hoq,Yoo:2022erv, Giesler:2022inPrep}. These new NR simulations provide valuable data that can be compared with results obtained from perturbative techniques such as the ppBHPT framework (including the \texttt{BHPTNRSur1dq1e4} surrogate model) and the second-order self-force model.

Building upon these recent advances, in this paper, we provide a detailed comparison between state-of-the-art NR simulations and perturbative results in the intermediate mass ratio regime.
We begin by providing an executive summary of the waveform data obtained from NR and point particle black hole perturbation theory (ppBHPT) in Section \ref{sec:waveforms}. In Section \ref{sec:waveform_comparison}, we conduct a comprehensive comparison of the dominant $(\ell,m) = (2,2)$ mode of the waveforms. We examine the phenomenology of the amplitudes and frequencies of different modes in Section \ref{sec:amp_freq} and discuss the differences in peak times of various spherical harmonic modes of the gravitational radiation in Section \ref{sec:peak_times}. 
To understand the non-linearities during the merger stage, we analyze the amplitude differences between NR and ppBHPT waveforms and propose fitting functions to describe these differences in Section \ref{sec:amplitude_differences}. Additionally, we evaluate the effectiveness of the $\alpha$-$\beta$ scaling technique in the intermediate mass ratio regime. We provide similar fits for the frequency differences in Section \ref{sec:freq_differences}.
Finally, in Section \ref{sec:discussion}, we discuss the implications and lessons learned for both NR and perturbative techniques.

%==========================================================================
%==========================================================================
%==========================================================================
\section{Gravitational waveforms in the intermediate mass ratio regime}
\label{sec:waveforms}
Gravitational radiation from the merger of a binary black hole is typically written as a superposition of $-2$ spin-weighted spherical harmonic modes with indices $(\ell,m$):
\begin{align}
h(t,\theta,\phi;\boldsymbol{\lambda}) &= \sum_{\ell=2}^\infty \sum_{m=-\ell}^{\ell} h^{\ell m}(t;\boldsymbol\lambda) \; _{-2}Y_{\ell m}(\theta,\phi)\,,
\label{hmodes}
\end{align}
where $\boldsymbol{\lambda}$ is the set of intrinsic parameters (such as the masses and spins of the binary) describing the system, $\theta$ is the polar angle, and $\phi$ is the azimuthal angle. In this paper, $h(t,\theta,\phi;\boldsymbol{\lambda})$ is obtained from both NR simulations and different flavors of perturbation theory frameworks.

%==========================================================================
\paragraph{\textbf{Numerical relativity data :}} 
We utilize the latest NR simulations of high mass ratio binaries performed by the RIT group ~\cite{Lousto:2022hoq,Lousto:2020tnb}. These simulations encompass mass ratios up to $q \le 128$ and spins ranging from $-0.85$ to $0.85$. The NR waveforms obtained from these simulations include modes up to $\ell=6$. However, due to numerical noise, we restrict our analysis to modes up to $\ell=4$ only. Additionally, for the current study, we focus exclusively on non-spinning cases.

%==========================================================================
\paragraph{\textbf{Perturbation theory waveforms :}}
We generate ppBHPT waveforms using the \texttt{BHPTNRSur1dq1e4} model~\cite{Islam:2022laz}, a recently developed surrogate waveform model that combines numerical relativity (NR) information with perturbation theory. This model can be accessed through the \texttt{gwsurrogate}\cite{gwsurrogate} or the \texttt{BHPTNRSurrogate}\cite{BHPTSurrogate} package from the \texttt{Black Hole Perturbation Theory Toolkit}~\cite{BHPToolkit}.

The \texttt{BHPTNRSur1dq1e4} model is trained on waveform data generated by the ppBHPT framework for non-spinning binaries with mass ratios ranging from $q=2.5$ to $q=10^{4}$. The full inspiral-merger-ringdown (IMR) ppBHPT waveform training data is computed using a time-domain Teukolsky equation solver, which has been extensively described in the literature~\cite{Sundararajan:2007jg,Sundararajan:2008zm,Sundararajan:2010sr,Zenginoglu:2011zz,Islam:2022laz,Rifat:2019ltp}. The model includes a total of 50 spherical harmonic modes up to $\ell=10$.

The model calibrates ppBHPT waveforms to NR data in the comparable mass ratio regime ($2.5 \le q \le 10$) up to $\ell=5$ employing a simple but non-trivial scaling called the $\alpha$-$\beta$ scaling~\cite{Islam:2022laz}:
\begin{align} \label{eq:EMRI_rescale}
h^{\ell,m}_{\tt full, \alpha_{\ell}, \beta}(t ; q) \sim {\alpha_{\ell}} h^{\ell,m}_{\tt pp}\left( t \beta;q \right) \,,
\end{align}
where $\alpha_{\ell}$ and $\beta$ are determined by minimizing the $L_2$-norm between the NR and rescaled ppBHPT waveforms. After this $\alpha$-$\beta$ calibration step, the ppBHPT waveforms exhibit remarkable agreement with NR waveforms (with an error of $\sim 10^{-3}$ for the $(2,2$ mode)). For instance, when compared to recent SXS and RIT NR simulations with mass ratios ranging from $q=15$ to $q=32$, the dominant quadrupolar mode of \texttt{BHPTNRSur1dq1e4} agrees to NR with errors smaller than $\approx 10^{-3}$.

Using \texttt{BHPTNRSurrogate}~\cite{BHPTSurrogate}, we then generate both ppBHPT and rescaled ppBHPT waveforms for any mass ratio within the training range of the model.

\begin{figure*}
\includegraphics[scale=0.52]{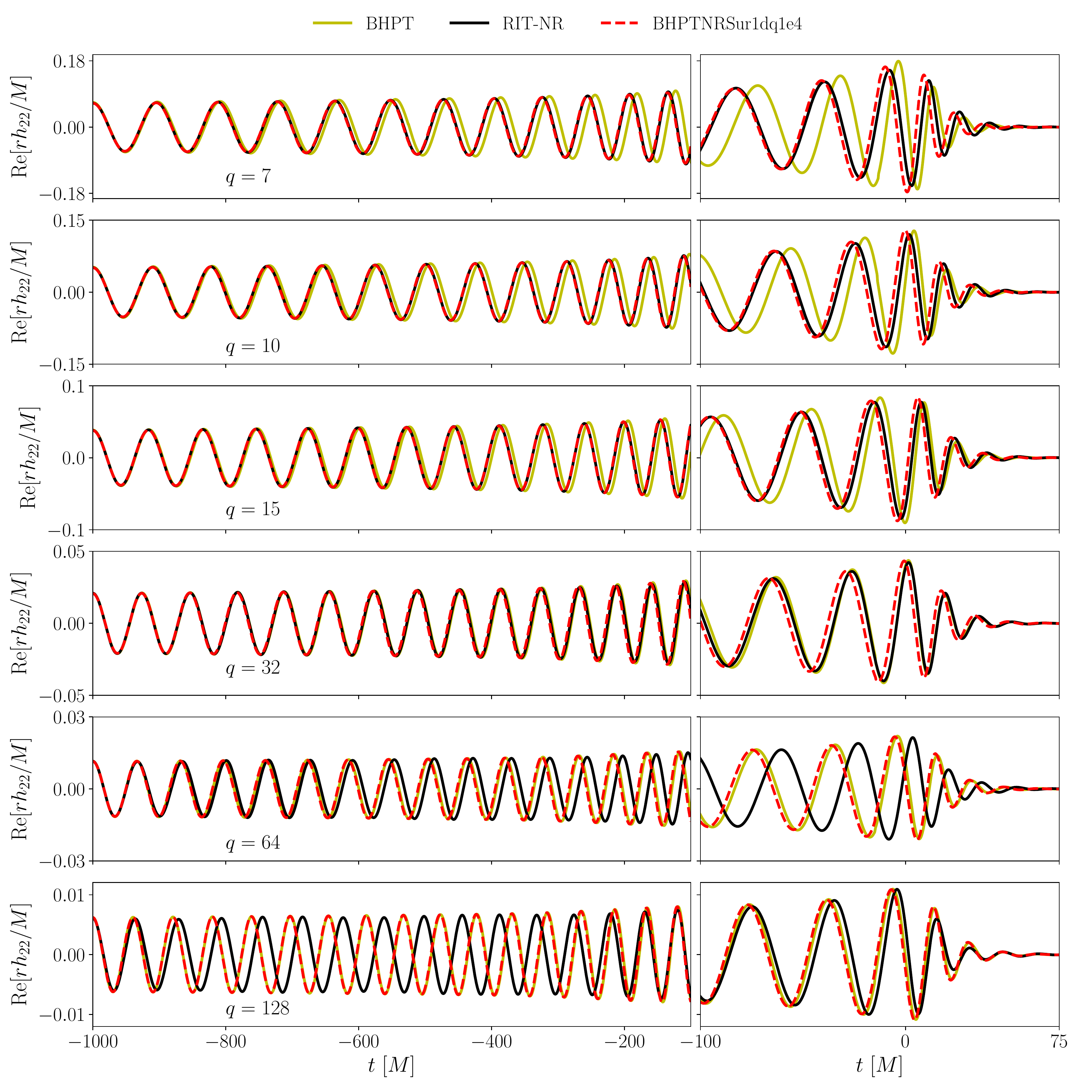}
\caption{We show the real part of the $(2,2)$ mode extracted from the NR data (solid black lines; labeled as `RIT-NR'), along with the ppBHPT waveform (solid yellow lines; labeled as `BHPT') and rescaled ppBHPT waveforms generated using the \texttt{BHPTNRSur1dq1e4} model (dashed red lines; labeled as `BHPTNRSur1dq1e4') for mass ratios $q=[7,15,32,64,128]$. More details are in Section \ref{sec:waveform_comparison}.}
\label{fig:waveforms}
\end{figure*}

%==========================================================================
%==========================================================================
%==========================================================================
\section{Comparison between NR and perturbation waveforms}
\label{sec:results}
Currently available high mass ratio NR simulations are of varying lengths, often spanning only $1500M$ (where $M$ is the total mass of the binary). This limited duration frequently poses a challenge when conducting a detailed comparison with existing waveform models. Additionally, many of the high mass ratio simulations exhibit residual eccentricity (see Appendix~\ref{app:ecc_RITNR}), further complicating waveform-level comparisons. Nonetheless, in Ref.~\cite{Islam:2022laz}, an interesting comparison is presented between RIT NR data and the \texttt{BHPTNRSur1dq1e4} waveform model for mass ratios $q=[15,32]$. While a comprehensive comparison of the full inspiral-merger-ringdown waveform is challenging due to the residual eccentricity in these simulations, they can still be utilized to comprehend and compare waveform phenomenology during the merger-ringdown stage, where the binary significantly circularizes. Hence, this paper primarily focuses on comparing the phenomenology of the NR data with the waveforms obtained from perturbation theory models.

%==========================================================================
%==========================================================================
\subsection{Comparison of $(\ell,m)=(2,2)$ mode waveforms}
\label{sec:waveform_comparison}
To begin, we decompose each spherical harmonics mode $h^{\ell m}(t)$ into its amplitude $A^{\ell m}(t)$ and phase $\phi^{\ell m}$ components, represented as $h^{\ell m}(t) = A^{\ell m}(t) e^{i\phi^{\ell m}}$.

For simplicity, we first focus on comparing the dominant $(\ell,m)=(2,2)$ mode during the final $\sim 1000M$ of the binary's evolution (see Fig.~\ref{fig:waveforms}). To facilitate this comparison, we align the multi-modal NR data (shown as solid black lines; labelled as `RIT-NR'), ppBHPT waveforms (shown as solid yellow lines; labelled as `BHPT'), and rescaled ppBHPT waveforms (represented by dashed red lines; labelled as `BHPTNRSur1dq1e4') on the same time grid $t=[-1000,100]M$, where $t=0M$ corresponds to the peak of the $(\ell,m)=(2,2)$ mode amplitude. Additionally, we adjust the phases such that the orbital phase is zero at the beginning of the waveforms i.e. at $t=-1000M$.

\begin{figure}
\includegraphics[width=\columnwidth]{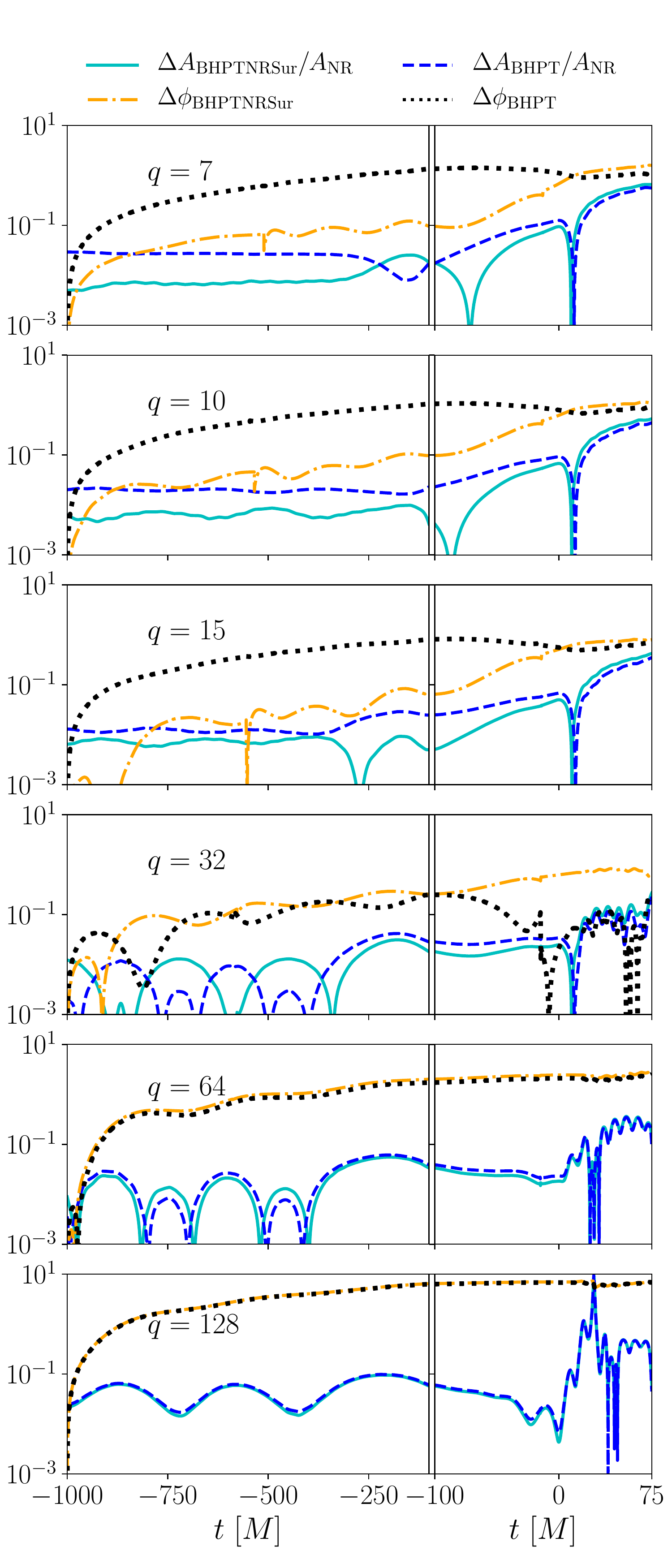}
\caption{We show the relative differences in amplitude $\Delta A_{22}/A_{22}^{\rm NR}$ and the absolute differences in phase $\Delta \phi_{22}$ for both ppBHPT and rescaled ppBHPT waveforms compared to the NR data for mass ratios $q=[7,15,32,64,128]$. More details are in Section \ref{sec:waveform_comparison}.}
\label{fig:errors}
\end{figure}

We observe that the rescaled ppBHPT waveforms exhibit a close match to the NR data for mass ratios ranging from $q=7$ to $q=32$, while the ppBHPT waveforms display differences in both amplitude and phase evolution when compared to NR data (top four rows of Fig.~\ref{fig:waveforms}). However, for mass ratios $q\geq64$, the NR data shows notable eccentricities, resulting in significant dephasing between the ppBHPT waveforms and NR, as well as between the rescaled ppBHPT waveforms and NR (bottom three rows of Fig.~\ref{fig:waveforms}).
Furthermore, it is important to mention that the ppBHPT and rescaled ppBHPT waveforms become increasingly similar to each other for mass ratios $q\geq64$. This suggests that the higher-order corrections to the linear ppBHPT results are relatively small in this regime.

\begin{figure*}
\includegraphics[width=\textwidth]{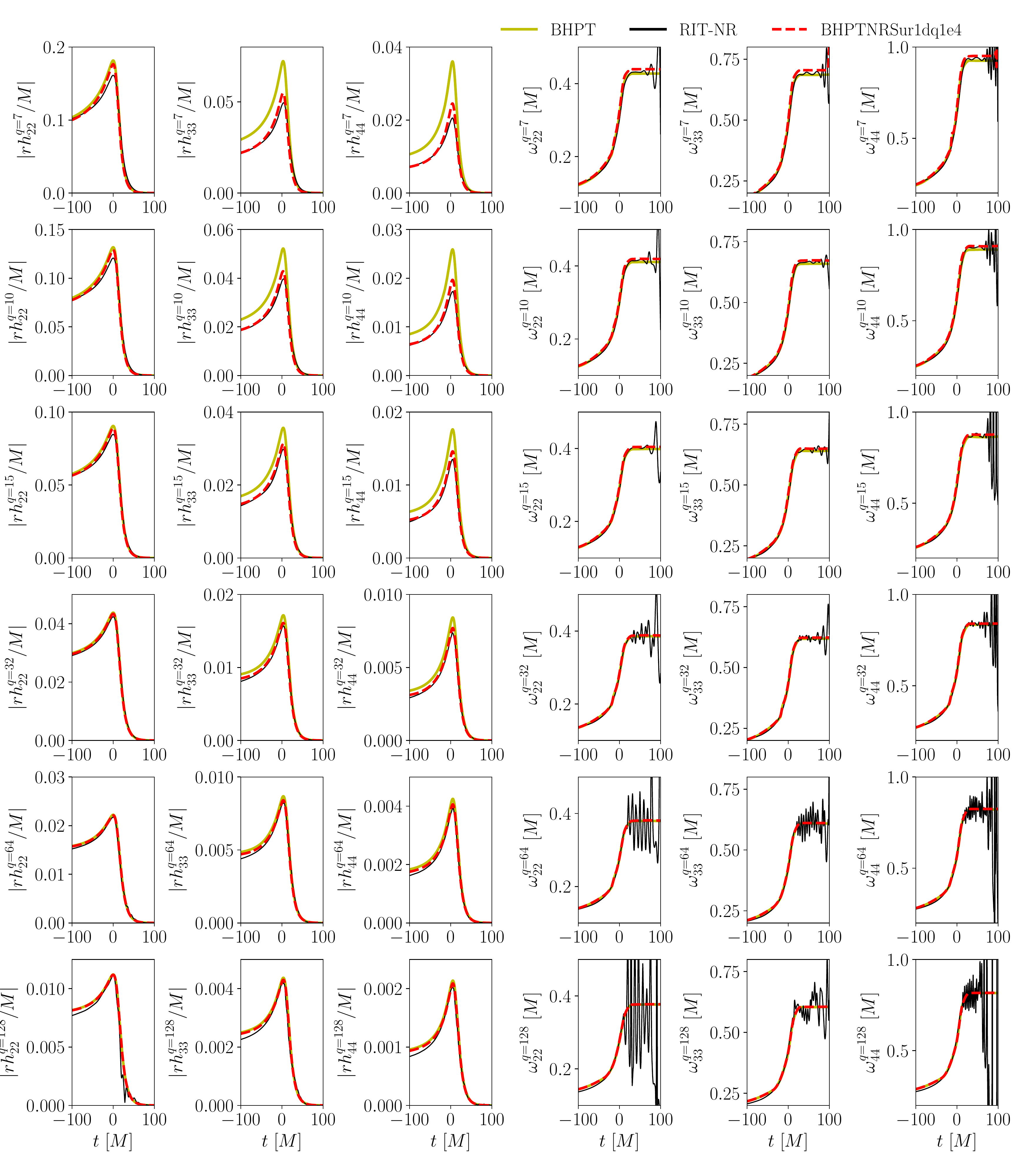}
\caption{We show the amplitudes and instantaneous frequencies of the $(2,2)$, $(3,3)$, and $(4,4)$ spherical harmonic modes extracted from the NR data (solid black lines), along with the amplitudes and instantaneous frequencies obtained from the ppBHPT waveforms (solid yellow lines; labeled as `BHPT') and rescaled ppBHPT waveforms generated using the \texttt{BHPTNRSur1dq1e4} model (dashed red lines; labeled as `BHPTNRSur1dq1e4') for mass ratios $q=[7,15,32,64,128]$. More details are in Section \ref{sec:waveform_comparison}.}
\label{fig:amp_freq}
\end{figure*}

In order to analyze the discrepancies between these waveforms, we calculate the relative differences in amplitude $\Delta A_{22}/A_{22}^{\rm NR}$ and the absolute differences in phase $\Delta \phi_{22}$ for both ppBHPT and rescaled ppBHPT waveforms compared to the NR data. Figure~\ref{fig:errors} illustrates the errors in amplitudes and phases during the late inspiral-merger-ringdown phase of the waveforms.
For mass ratios in the range of $q=7$ to $q=16$, it is clear that the differences in both amplitudes and phases between the rescaled ppBHPT waveforms and the NR waveforms are significantly smaller than those observed between the ppBHPT waveforms and NR. This suggests that the linear ppBHPT waveforms are insufficient in accurately matching the NR waveforms within this mass ratio range.
However, as we move towards higher mass ratios (i.e. $q\geq32$), the differences in $\Delta A_{22}/A_{22}^{\rm NR}$ and $\Delta \phi_{22}$ between the ppBHPT and rescaled ppBHPT waveforms diminish gradually. This indicates that the linear description of the binary evolution becomes increasingly accurate as the mass ratio increases.
For mass ratios $q\geq64$, both $\Delta A_{22}/A_{22}^{\rm NR}$ and $\Delta \phi_{22}$ exhibit distinct features that strongly suggest the presence of residual eccentricities in the NR simulations.

At this point, we aim to quantify the difference between NR and (scaled) ppBHPT waveforms for different mass ratios using the $L_2$-norm. To compute the $L_2$-norm between two waveforms $h_1^{22}(t)$ and $h_2^{22}(t)$, we minimize the time-domain overlap integral (or $L_2$-norm error) given by:
\begin{align} \label{eq:old_opt}
{\cal E} (h_1, h_2)=\min_{t_c,\varphi_z} \frac{\int \left| \mathrm{e}^{- 2 \pi \mathrm{i} \varphi_z}h_1^{22}(t - t_c) - h_2^{22}(t) \right|^2 dt}{\int \left| h^{22}_2(t) \right|^2 dt} ,.
\end{align}
We compute this error over a shift in time $t_c$ and a rotation about the z-axis by an angle $\varphi_z$. It is important to note that the duration of NR simulations varies significantly for different mass ratios. For instance, at $q=7$, the NR data covers the final $\sim2400M$ of the binary evolution, while for $q=64$, the NR data is only $\sim 1000M$ long. Initially, we use all available NR data to compute these differences. We find that the scaled ppBHPT waveforms (obtained from \texttt{BHPTNRSur1dq1e4}) yield a better match to NR than the original ppBHPT waveforms. For example, the $L_2$-norm error between NR and ppBHPT waveforms is $\sim0.6$, while the error between NR and scaled ppBHPT waveforms is $\sim0.03$ for $q=10$. However, for $q=64$, both ppBHPT and scaled ppBHPT waveforms exhibit equally worse agreement with NR data (which includes residual eccentricity), yielding $L_2$-norm values of $0.6$ and $0.7$, respectively. Similar errors are also obtained for $q=128$. We further point out that the errors are worse for the higher order modes.

\subsection{Comparison of the amplitudes and frequencies of different modes}
\label{sec:amp_freq}
We now examine the amplitudes and instantaneous frequencies of three representative modes $[(\ell,m)]=[(2,2),(3,3),(4,4)]$ for mass ratios ranging from $q=7$ to $q=128$ (see Fig.~\ref{fig:amp_freq}). For any given mode, instantaneous frequencies $\omega_{\ell,m}$ is given by the time derivative of the phase
\begin{equation}
    \omega_{\ell,m} = \frac{d\phi_{\ell,m}}{dt}.
\end{equation}
To mitigate the impact of residual eccentricities in the comparisons, we focus on the merger-ringdown stage of the binary ($-100M \le t \le 100M$), where circularization is expected to be nearly complete.
For mass ratios $7 \le q \le 32$, noticeable differences are observed between ppBHPT and NR amplitudes, while the rescaled ppBHPT amplitudes closely match the NR values across all mass ratios. Moreover, as anticipated, the differences in amplitudes between ppBHPT and NR (and rescaled ppBHPT) decrease as the mass ratio increases. For $q\ge 64$, ppBHPT and rescaled ppBHPT produce nearly identical amplitudes.
Interestingly, the frequencies of the individual modes computed from ppBHPT waveforms and NR exhibit remarkable agreements for all mass ratios. It is important to note that due to numerical noise in the NR data, frequencies display unphysical oscillations after the merger, particularly for mass ratios $q\ge15$.

\begin{figure}
\includegraphics[width=\columnwidth]{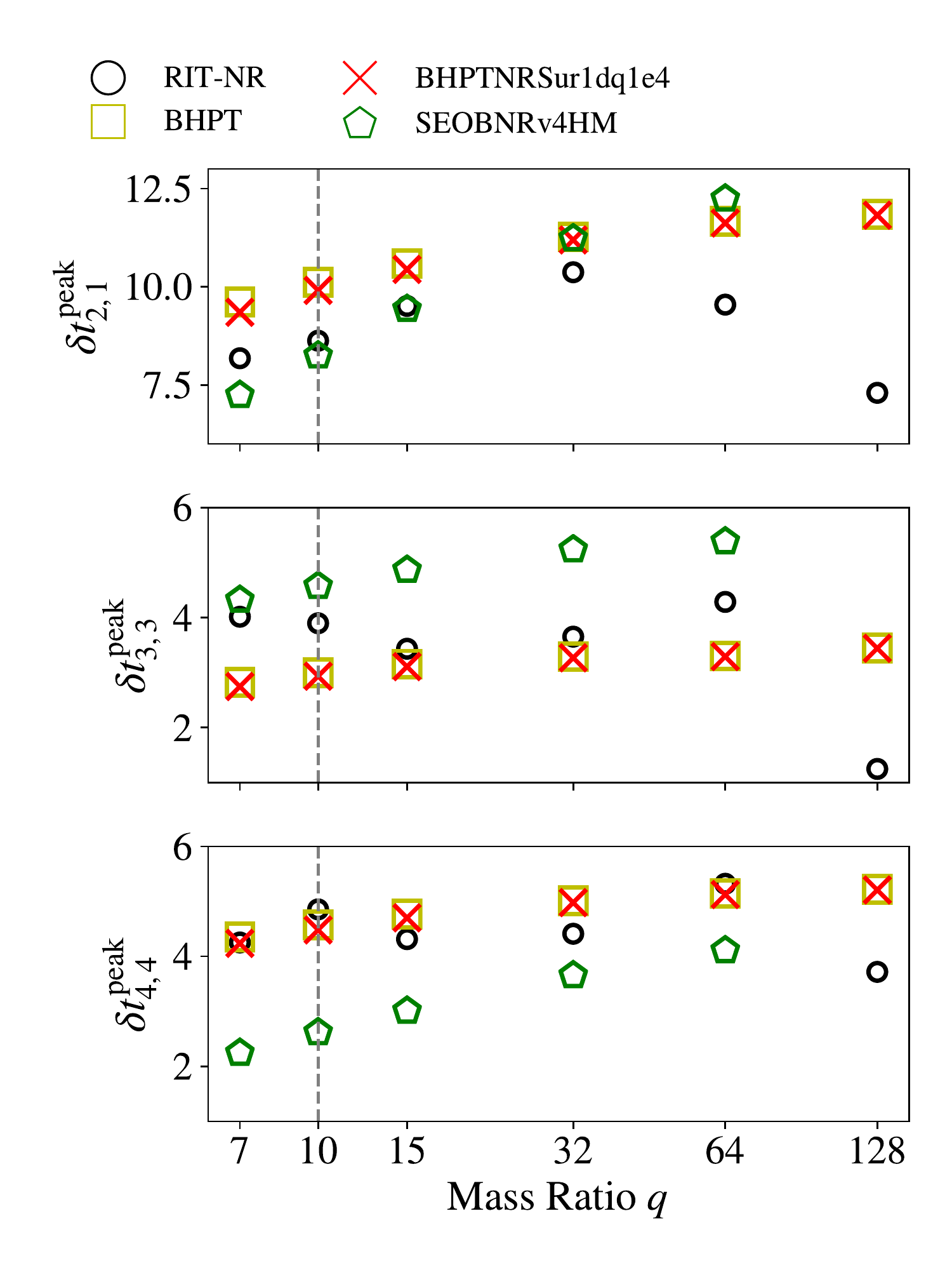}
\caption{We show the relative times (with respect to the $(2,2)$ mode) at which the amplitudes of the $(2,1)$ mode (upper panel), $(3,3)$ mode (middle panel), and $(4,4)$ mode (lower panel) reach the maximum in the NR data (circles) along with the times obtained from the ppBHPT waveforms (squares; labeled as `BHPT') and rescaled ppBHPT waveforms generated using the \texttt{BHPTNRSur1dq1e4} model (crosses; labeled as `BHPTNRSur1dq1e4'). Additionally, we include the relative peak locations in the \texttt{SEOBNRv4HM} model (represented by pentagons) for comparison. The grey vertical dashed line represents $q=10$, which serves as a crude boundary between the comparable mass regime and the intermediate mass ratio regime. More details are in Section~\ref{sec:peak_times}.}
\label{fig:nonspinning_peak_time}
\end{figure}

%==========================================================================
%==========================================================================
\subsection{Comparison of the peak times}
\label{sec:peak_times}
Next, we determine the times $t_{\ell,m}^{\rm peak}$ corresponding to the maximum amplitude $A^{\rm peak}{\ell,m}$ for each spherical harmonic mode. We then calculate the relative time of the peaks with respect to the dominant $(2,2)$ mode as:
\begin{equation}
\delta t^{\rm peak}{\ell,m} = t_{\ell,m}^{\rm peak} - t_{2,2}^{\rm peak},
\end{equation}
where $t_{2,2}^{\rm peak}$ is the time at which the $(2,2)$ mode amplitude reaches its maximum. We show the relative peak times $\delta t^{\rm peak}_{\ell,m}$ in the NR data for a set of three representative modes $[(\ell,m)]=[(2,1),(3,3),(4,4)]$ along with the relative peak times for the same modes in the ppBHPT and rescaled ppBHPT waveforms in Fig.~\ref{fig:nonspinning_peak_time}.
For comparison, we include the relative peak times of these modes from one of the state-of-the-art effective-one-body models for aligned-spin binaries, namely \texttt{SEOBNRv4HM}. This model includes four higher-order modes in addition to the dominant quadrupolar mode of radiation: $(\ell,m)=[(2,\pm1),(3,\pm3),(4,\pm4),(5,\pm5)]$, and it is calibrated to a set of 141 NR waveforms for mass ratios $q\le10$ and spins $\chi_{1,2} \le 0.99$.

\begin{figure}
\includegraphics[width=\columnwidth]{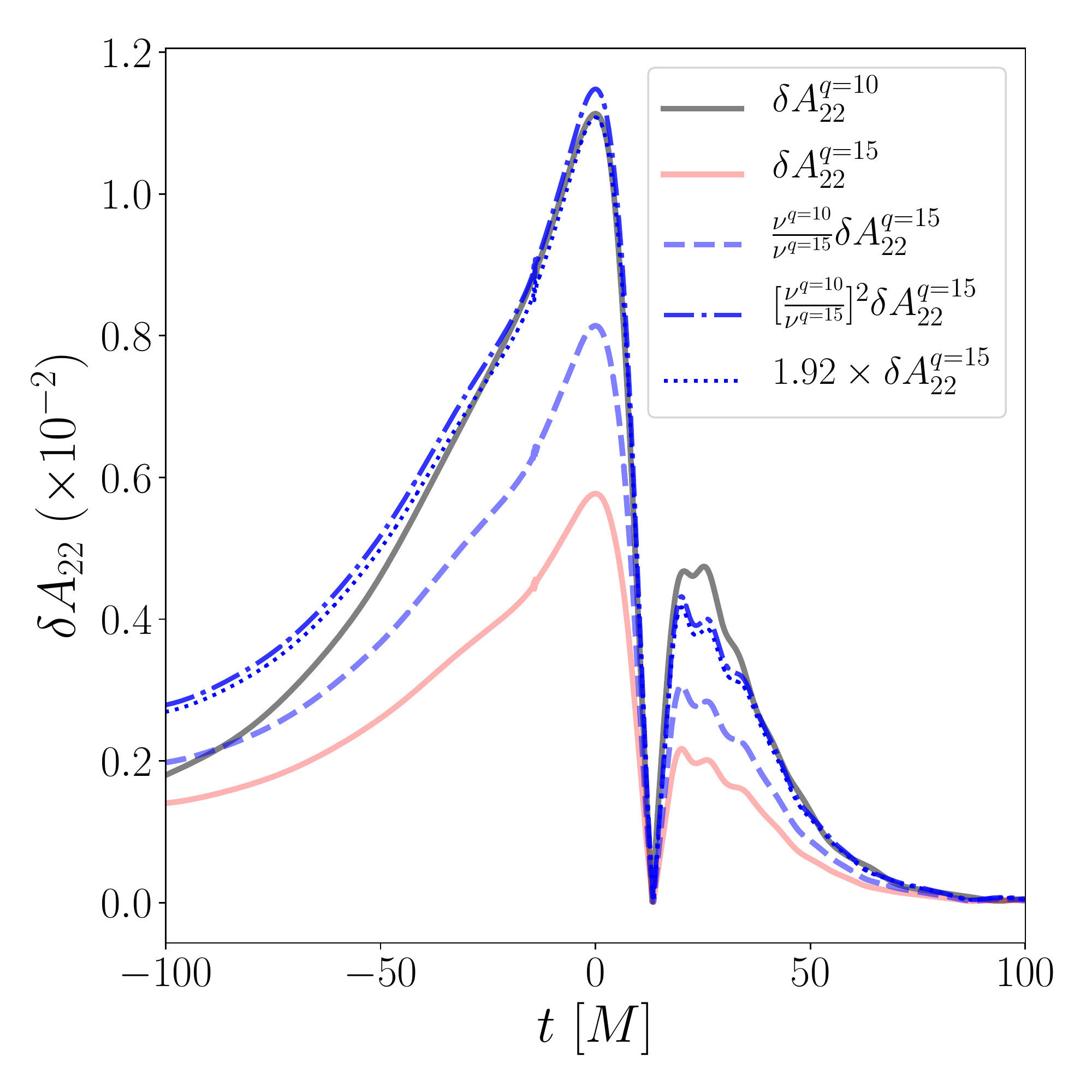}
\caption{We show the amplitude difference $\delta A_{22}$ between ppBHPT and NR for $q=10$ (solid grey line) and $q=15$ (sloid red line). Furthermore, we show the amplitude difference for $q=15$ after rescaling them with different powers of the ratio of symmetric mass ratios (blue lines). More details are in Section~\ref{sec:amplitude_differences}.}
\label{fig:q10q15_deltaA}
\end{figure}

Interestingly, the relative peak times $\delta t^{\rm peak}_{\ell,m}$ within these waveforms exhibit significant inconsistencies with each other for almost all mass ratio values.
The inconsistencies in the relative peak times $\delta t^{\rm peak}_{\ell,m}$ indicate that there is still room for improvement in accurately predicting the timing of different modes during the merger-ringdown phases of binary black hole systems. Further developments in waveform modeling techniques and more comprehensive calibration against NR simulations may help reduce the discrepancies. 
We further notice that the differences in peak times between ppBHPT and rescaled ppBHPT waveforms are very small. This can be attributed to the dominant influence of the inspiral phase in the $\alpha$-$\beta$ calibration procedure. Accurate modelling of the peak times in rescaled ppBHPT waveforms (i.e. in \texttt{BHPTNRSur1dq1e4}) may require further tuning in the merger-ringdown part as done in Ref.~\cite{Islam:2023mob}.

\begin{figure}
\includegraphics[width=\columnwidth]{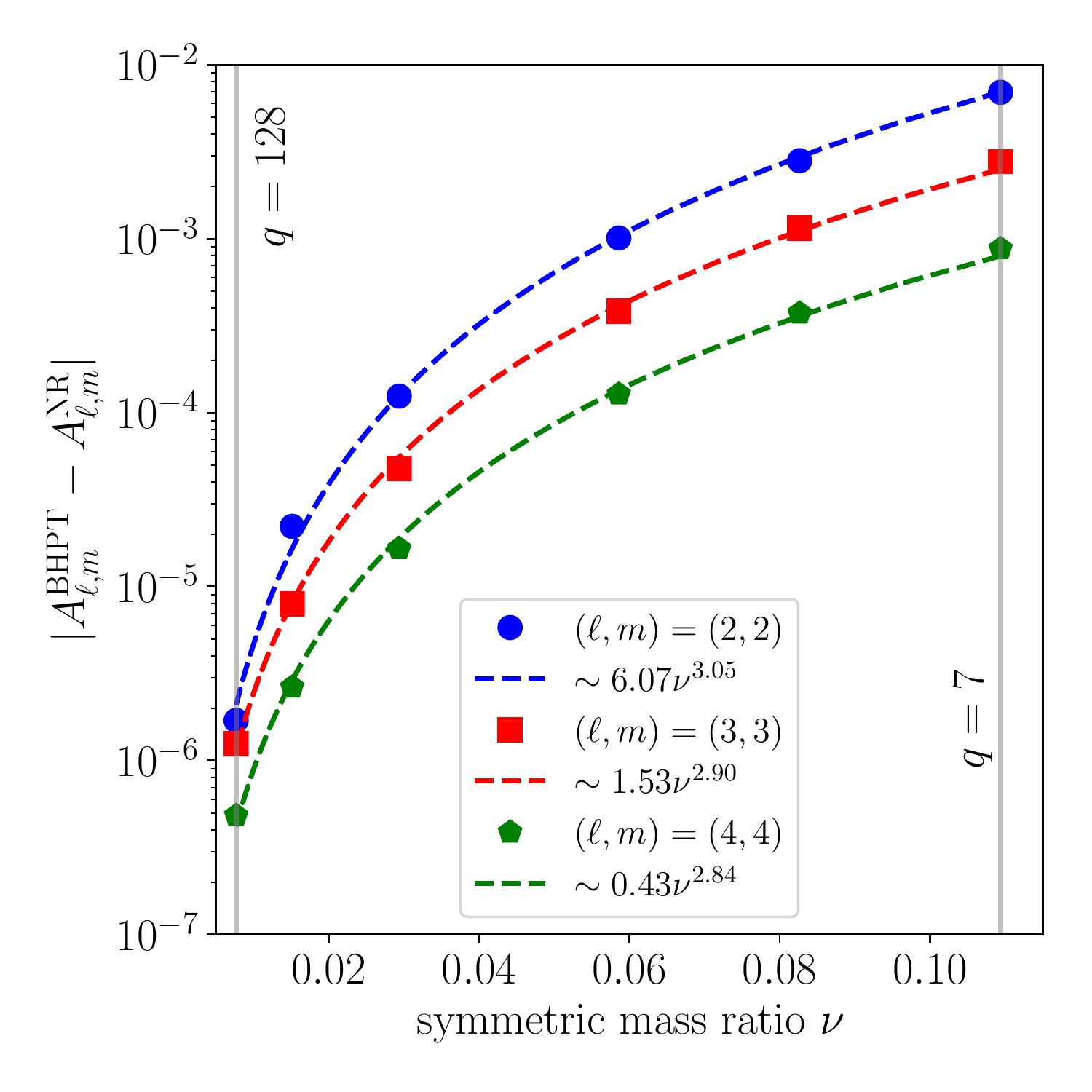}
\caption{We show the amplitude differences $\delta A_{\ell,m}$ between ppBHPT and NR data at the merger (denoted by the maximum amplitude in the $(2,2)$ mode) for $(2,2)$ (blue circles), $(3,3)$ (red squares) and $(4,4)$ (green pentagons) modes for mass ratios $q=[7,15,32,64,128]$. Additionally, we show the best-fit functions for each mode in terms of $\nu$. More details are in Section \ref{sec:amplitude_differences}.}
\label{fig:deltaA_nu_fits}
\end{figure}

%==========================================================================
%==========================================================================
%==========================================================================
%==========================================================================
\section{Interplay between NR and perturbation theory}
\label{sec:interplay}
To gain a deeper understanding of the interaction between the NR and ppBHPT waveforms, we now examine their disparities in terms of amplitude and frequencies (as illustrated in Figure~\ref{fig:amp_freq}) across different mass ratios. 

%==========================================================================
%==========================================================================
\subsection{Amplitude differences between NR and ppBHPT}
\label{sec:amplitude_differences}
We first investigate the differences between NR and ppBHPT in amplitude across various mass ratios. Specifically, we replicate and expand upon the analysis presented in Refs.~\cite{Lousto:2010tb,Lousto:2010qx,Nakano:2011pb}. Following the methodology outlined in Refs.~\cite{Lousto:2010tb,Lousto:2010qx,Nakano:2011pb}, we define the amplitude differences as:
\begin{equation}
\delta A_{\ell,m} = | A_{\ell,m}^{\text{BHPT}} - A_{\ell,m}^{\text{NR}} |,
\end{equation}
where $A_{\ell,m}^{\text{BHPT}}$ represents the amplitude of the ppBHPT waveform.

\begin{figure}
\includegraphics[width=\columnwidth]{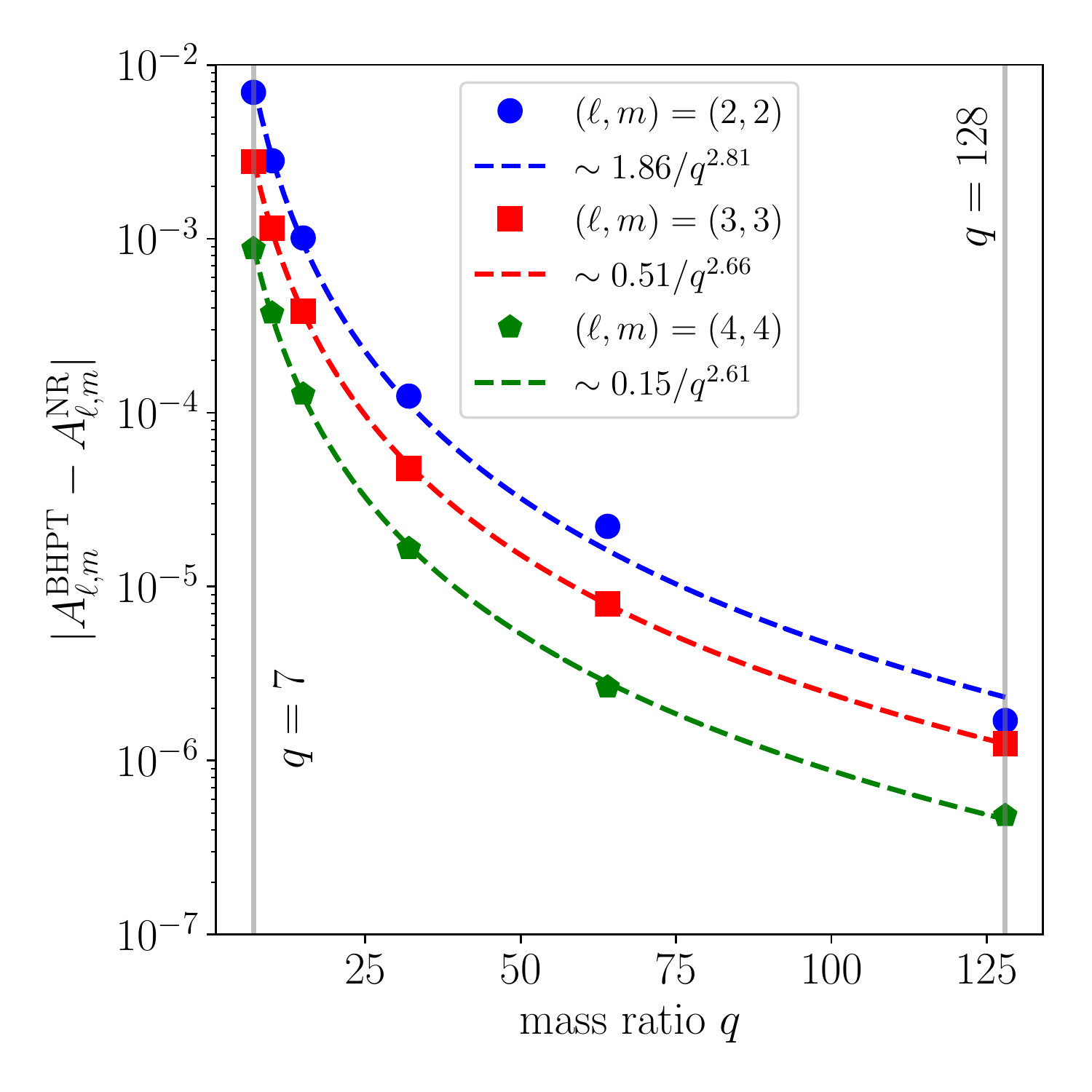}
\caption{We show the amplitude differences $\delta A_{\ell,m}$ between ppBHPT and NR data at the merger (denoted by the maximum amplitude in the $(2,2)$ mode) for $(2,2)$ (blue circles), $(3,3)$ (red squares) and $(4,4)$ (green pentagons) modes for mass ratios $q=[7,15,32,64,128]$. Additionally, we show the best-fit functions for each mode in terms of $\frac{1}{q}$. More details are in Section \ref{sec:amplitude_differences}.}
\label{fig:deltaA_smallq_fits}
\end{figure}

We observe that the amplitude differences for the $q=10$ and $q=15$ cases near the merger exhibit the following behavior (Fig.~\ref{fig:q10q15_deltaA}):
\begin{align}
\delta A_{22}^{q=10} &\sim 1.92 \times \delta A_{22}^{q=15}\notag\\
&\sim 1.44^{1.92} \times \delta A_{22}^{q=15},
\end{align}
where 1.44 is to the ratio of the symmetric mass ratios $\nu$. This approximate scaling differs slightly from the one reported in Ref.~\cite{Nakano:2011pb}, which suggested $\delta A_{22}^{q=10} \sim 1.44^{2.3} \times \delta A_{22}^{q=15}$. Nevertheless, both results indicate the presence of nonlinear effects (beyond adiabatic evolution) in the amplitude differences between the NR and ppBHPT waveforms, as these differences scale nonlinearly with the symmetric mass ratio $\nu$.
Likewise, we find that the amplitude differences for the $q=10$ and $q=32$ cases near the merger can be characterized as follows:
\begin{align}
\delta A_{22}^{q=10} &\sim 7.7 \times \delta A_{22}^{q=32}\notag\\
&\sim 2.81^{1.98} \times \delta A_{22}^{q=32},
\end{align}
where 2.81 is the ratio of the symmetric mass ratios $\nu$.
Similarly, the amplitude differences for the $q=15$ and $q=32$ cases near the merger obeys:
\begin{align}
\delta A_{22}^{q=10} &\sim 3.97 \times \delta A_{22}^{q=32}\notag\\
&\sim 1.99^{1.96} \times \delta A_{22}^{q=32},
\end{align}
where 1.99 is the ratio of the symmetric mass ratios $\nu$.

Next, we perform fitting for the amplitude differences $\delta A_{\ell,m}$ of three representative modes $(\ell,m)=[(2,2),(3,3),(4,4)]$ at their respective peaks as a function of $\nu$ (Fig.~\ref{fig:deltaA_nu_fits}). The obtained relations are as follows:
\begin{align}
\delta A_{2,2} \sim 6.07 \times \nu^{3.06}\\
\delta A_{3,3} \sim 1.53 \times \nu^{2.90}\\
\delta A_{4,4} \sim 0.43 \times \nu^{2.84}.
\end{align}
Next, we repeat the fitting in terms of $\frac{1}{q}$ ((Fig.~\ref{fig:deltaA_smallq_fits})) and find:
\begin{align}
\delta A_{2,2} \sim 1.86 / q^{2.81}\\
\delta A_{3,3} \sim 0.51 / q^{2.66}\\
\delta A_{4,4} \sim 0.15 / q^{2.61}.
\end{align}
These fits not only provide a simple scaling for the differences in maximum amplitudes between ppBHPT and NR waveforms, but also serve as further confirmation of the presence of non-linearity during the merger stage of the binary evolution. Additionally, we observe that the non-linearity is more pronounced in the $(2,2)$ mode compared to higher order modes.

\begin{figure}
\includegraphics[width=\columnwidth]{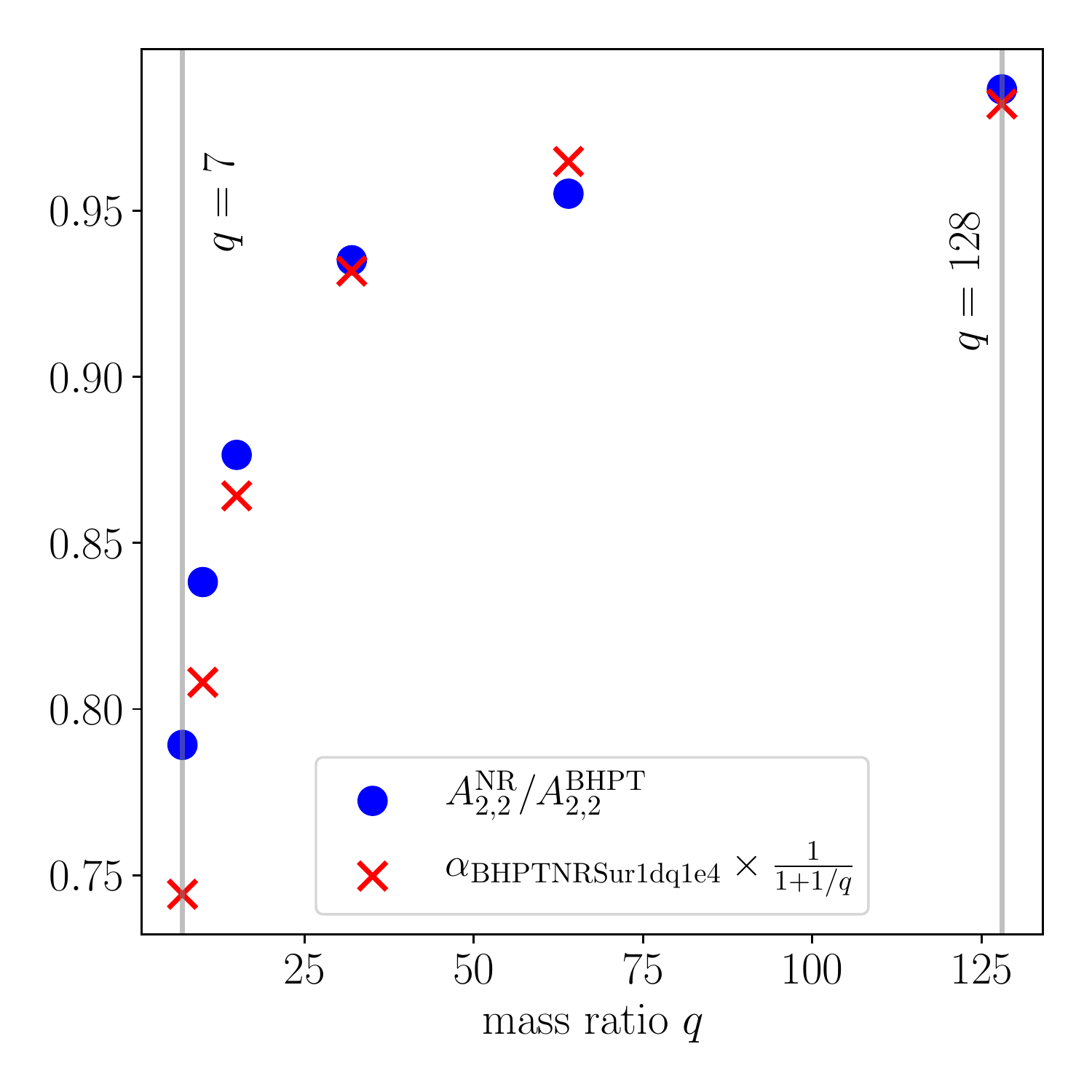}
\caption{We show the ratio of the ppBHPT and NR amplitudes for mass ratios $q=[7,15,32,64,128]$, along with the $\alpha$ parameter extracted from the \texttt{BHPTNRSur1dq1e4} model. More details are in Section \ref{sec:amplitude_differences}.}
\label{fig:alpha}
\end{figure}

We now calculate $A_{\rm NR}/A_{\rm BHPT}$, which represents the ratio of the ppBHPT and NR amplitudes for all mass ratios. This ratio is expected to correspond roughly to the $\alpha$ parameter in Eq.~(\ref{eq:EMRI_rescale}) after multiplying by the transformation factor $\frac{1}{1+1/q}$ between a mass scale of $m_1$ and $M$. In Figure \ref{fig:alpha}, we present both the ratio of the amplitude $A_{\rm NR}/A_{\rm BHPT}$ and the $\alpha$ values obtained from the \texttt{BHPTNRSur1dq1e4} model. We observe that as the mass ratio increases, the agreement between these two quantities improves, suggesting that the $\alpha$-$\beta$ scaling works reasonably well even beyond the comparable mass ratio regime where it was originally constructed. The differences observed for $q\le 15$ can be attributed to numerical noise and the presence of residual eccentricities in the NR data.

Next, we repeat our study using scaled ppBHPT waveforms (obtained from the \texttt{BHPTNRSur1dq1e4} model). In particular, we calculate the differences in amplitude across various mass ratios. 
'\begin{equation}
\delta A^{\rm scaled}_{\ell,m} = | A_{\ell,m}^{\text{BHPTNRSur1dq1e4}} - A_{\ell,m}^{\text{NR}} |,
\end{equation}
where $A_{\ell,m}^{\text{BHPTNRSur1dq1e4}}$ represents the amplitude of the scaled ppBHPT waveform. At the peak, we find the following relations:
\begin{align}
\delta A^{\rm scaled}_{2,2} \sim 4.87 \times \nu^{3.09}\\
\delta A^{\rm scaled}_{3,3} \sim 0.13 \times \nu^{2.53}\\
\delta A^{\rm scaled}_{4,4} \sim 0.03 \times \nu^{2.46}
\end{align}
and
\begin{align}
\delta A^{\rm scaled}_{2,2} \sim 1.47 / q^{2.83}\\
\delta A^{\rm scaled}_{3,3} \sim 0.05 / q^{2.33}\\
\delta A^{\rm scaled}_{4,4} \sim 0.014 / q^{2.26}.
\end{align}
It is worth noting that the exponents in the relation for $\delta A^{\rm scaled}_{\ell,m}$ have changed only slightly compared to $\delta A_{\ell,m}$. However, it is important to highlight that the coefficients for $\delta A^{\rm scaled}_{\ell,m}$ are much smaller than the ones that appear in $\delta A_{\ell,m}$. 

%==========================================================================
%==========================================================================
\subsection{Frequency differences between NR and ppBHPT}
\label{sec:freq_differences}
Following the methodology described in Section \ref{sec:amplitude_differences} regarding the amplitudes, we define the frequency differences as:
\begin{equation}
\delta \omega_{\ell,m} = |\omega_{\ell,m}^{\text{BHPT}} - \omega_{\ell,m}^{\text{NR}}|,
\end{equation}
where $\omega_{\ell,m}^{\text{BHPT}}$ and $\omega_{\ell,m}^{\text{NR}}$ represent the instantaneous frequencies of the ppBHPT and NR waveforms, respectively.

\begin{figure}
\includegraphics[width=\columnwidth]{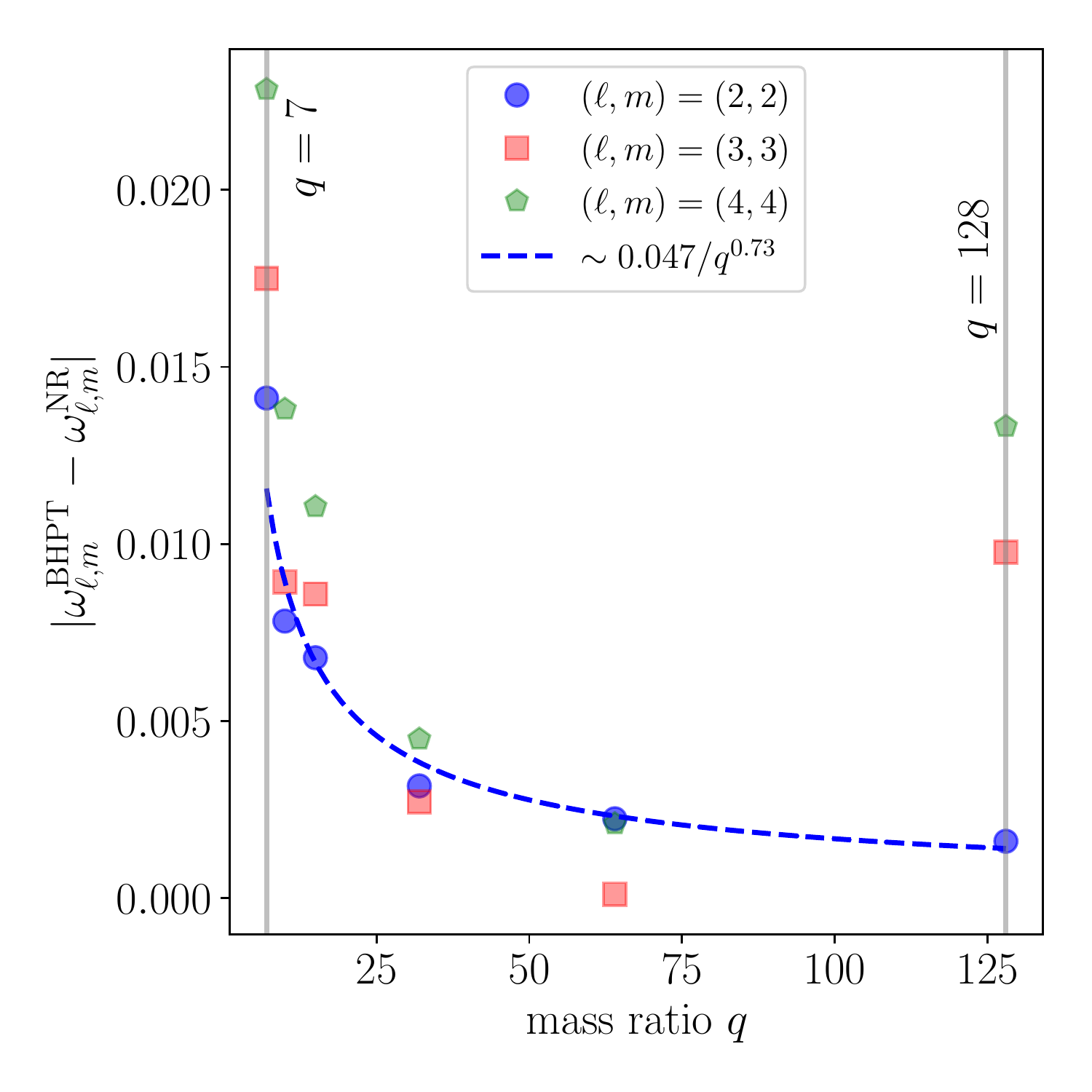}
\caption{We show the frequency differences $\delta \omega_{\ell,m}$ between ppBHPT and NR data at the merger (denoted by the maximum amplitude in the $(2,2)$ mode) for $(2,2)$ (blue circles), $(3,3)$ (red squares) and $(4,4)$ (green pentagons) modes for mass ratios $q=[7,15,32,64,128]$. Additionally, we show the best-fit functions for the $(2,2)$ mode in terms of $\frac{1}{q}$. More details are in Section \ref{sec:freq_differences}.}
\label{fig:deltaomega_smallq_fits}
\end{figure}

We calculate $\delta \omega_{\ell,m}$ at the merger, indicated by the maximum amplitude in the $(2,2)$ mode, for the $(2,2)$, $(3,3)$ and $(4,4)$ modes for mass ratios $q=[7,15,32,64,128]$ (Fig.~\ref{fig:deltaomega_smallq_fits}). Subsequently, we conduct a fitting analysis for the frequency differences $\delta A_{2,2}$ in terms of $\frac{1}{q}$ and obtain the following relationship (Fig.~\ref{fig:deltaomega_smallq_fits}):
\begin{equation}
\delta \omega_{2,2} \sim 0.047 / q^{0.73}.\\
\end{equation}
Next, we repeat the fitting in terms of $\nu$ and find:
\begin{align}
\delta \omega_{2,2} \sim 0.063 \nu^{0.78}.
\end{align}

It is important to acknowledge that due to numerical noise present in the NR data, as observed in Figure~\ref{fig:amp_freq}, it becomes increasingly difficult to obtain accurate estimates of the instantaneous frequencies from NR for mass ratios $q\geq16$. Therefore, we refrain from attempting to fit the frequency differences for the $(3,3)$ and $(4,4)$ modes in this scenario.

We then investigate the differences in the instantaneous frequencies between scaled ppBHPT and NR around merger, defined as:
\begin{equation}
\delta \omega^{\rm scaled}_{2,2} = |\omega_{2,2}^{\text{BHPTNRSur1dq1e4}} - \omega_{2,2}^{\text{NR}}|.
\end{equation}
We find the following scalings for $\delta \omega^{\rm scaled}_{2,2}$:
\begin{equation}
\delta \omega^{\rm scaled}_{2,2} \sim 0.233 / q^{1.07},
\end{equation}
and
\begin{align}
\delta \omega^{\rm scaled}_{2,2} \sim 0.154 \nu^{0.98}.
\end{align}
In contrast to the amplitude differences, we observe that both the coefficient and exponent are significantly different between $\delta \omega^{\rm scaled}_{2,2}$ and $\delta \omega_{2,2}$ scalings.

%==========================================================================
%==========================================================================
%==========================================================================
\section{Discussions \& Conclusion}
\label{sec:discussion}
In this work, we have conducted a detailed comparison between state-of-art NR simulations and perturbative results in the intermediate mass ratio regime. In particular, we use both ppBHPT waveforms and rescaled ppBHPT waveforms from the \texttt{BHPTNRSur1dq1e4} surrogate model.

We first provide a comprehensive comparison of the dominant $(\ell,m)=(2,2)$ mode  of the gravitational radiation obtained from NR and ppBHPT techniques. We observe that the rescaled ppBHPT waveforms exhibit a close match to the NR data for mass ratios ranging from $q=7$ to $q=32$, while the ppBHPT waveforms display differences in both amplitude and phase evolution when compared to NR data. For mass ratios $q\ge32$, residual eccentricities and numerical noise in the NR data make such comparisons challenging (Section \ref{sec:waveform_comparison}; Fig.~\ref{fig:waveforms} and Fig.~\ref{fig:errors}). We further observe that as the mass ratio increases, the differences between NR data and ppBHPT results reduce (Section \ref{sec:waveform_comparison}; Fig.~\ref{fig:amp_freq}). 
Furthermore, excellent match between NR amplitudes and scaled ppBHPT amplitudes indicate effectiveness of the $\alpha$-$\beta$ scaling in the intermediate mass ratio regime (Section \ref{sec:amplitude_differences}; Fig.~\ref{fig:alpha}).  
However, the differences in peak times of different modes between NR, ppBHPT and \texttt{BHPTNRSur1dq1e4} highlight the intricacies of the merger stage, revealing insights into the non-linear dynamics of the binary evolution (Section \ref{sec:peak_times}; Fig.~\ref{fig:nonspinning_peak_time}). 

Next, we examine the disparities between NR and ppBHPT waveforms in terms of amplitude and frequencies to gain a comprehensive understanding of the intricate relationship between these two frameworks. We analyze the amplitude differences $\delta A_{\ell,m}$ between NR and ppBHPT waveforms for different modes to investigate the non-linearities present during the merger stage of the binary evolution and propose fitting functions to describe these amplitude differences in terms of both $q$ and $\nu$. The proposed fitting functions for amplitude differences between NR and ppBHPT waveforms offer a valuable tool for understanding and quantifying these non-linearities (Section \ref{sec:amplitude_differences}; Figs.~\ref{fig:q10q15_deltaA},~\ref{fig:deltaA_nu_fits},~\ref{fig:deltaA_smallq_fits}). Finally, we provide similar fits for the frequency differences in the $(2,2)$ mode in Section~\ref{sec:freq_differences}.

This study highlights the potential of ppBHPT and surrogate models, such as \texttt{BHPTNRSur1dq1e4}, in efficiently and accurately predicting waveforms in the intermediate mass ratio regime. It opens up new opportunities for exploring the non-linearities during the merger stage of binary and for developing reliable modeling strategies to accurately determine the peak times of each mode. Our findings underscore the importance of improving calibration methods for ppBHPT-based surrogate models and enhancing eccentricity reduction algorithms in NR simulations. These advancements will contribute to the development of more accurate and efficient waveform models, enabling better detection and characterization of GW signals in the intermediate mass ratio regime.

%==========================================================================
%==========================================================================
%==========================================================================
\begin{acknowledgments}
T.I. would like to thank Gaurav Khanna and Scott Field for helpful discussion. T.I. is supported by NSF Grants No. PHY-1806665 and DMS-1912716. This work is performed on CARNiE at the Center for Scientific Computing and Visualization Research (CSCVR) of UMassD, which is supported by the ONR/DURIP Grant No.\ N00014181255, the UMass-URI UNITY supercomputer supported by the Massachusetts Green High Performance Computing Center (MGHPCC) and ORNL SUMMIT under allocation AST166. 
\end{acknowledgments}  

%==========================================================================
%==========================================================================
%==========================================================================
\appendix
\section{Residual eccentricities in RIT-NR simulations} 
\label{app:ecc_RITNR}

\begin{figure}
\includegraphics[width=\columnwidth]{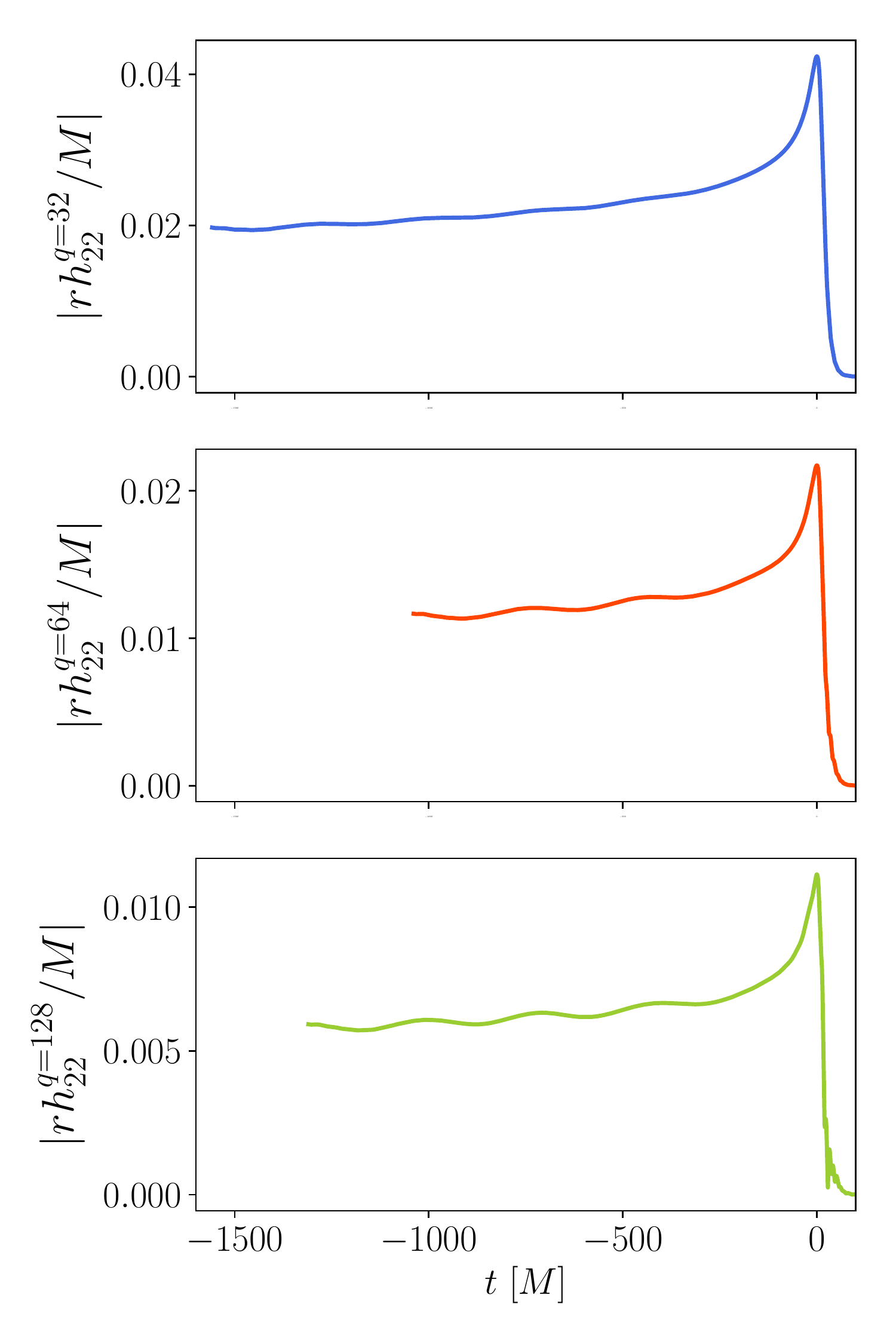}
\caption{We show the amplitudes of the $(2,2)$ mode in RIT-NR waveform data for three different mass ratios: $q=[32,64,128]$. In all cases, we see modulations in the amplitudes due to residual eccentricity. More details are in Appendix~\ref{app:ecc_RITNR}.}
\label{fig:ecc_amp}
\end{figure}

It is important to highlight the challenges associated with using RIT NR simulations~\cite{Lousto:2020tnb,Lousto:2022hoq} to estimate the accuracy of waveform models in the intermediate mass ratio regime. Two notable limitations are the shorter length of the NR data and the presence of residual eccentricities.

To illustrate these issues, we plot the amplitudes of the $(2,2)$ mode in RIT-NR waveform data in Fig.~\ref{fig:ecc_amp} for three different mass ratios: $q=32$ (\texttt{RIT-BBH-0792}; upper panel), $q=64$ (\texttt{RIT-BBH-1916}; middle panel), and $q=128$ (\texttt{RIT-BBH-1076}; lower panel). We have chosen the same time-range for all three subplots to stress the varying (and relatively short) length of the NR data corresponding to different mass ratios. Figure ~\ref{fig:ecc_amp} further shows clear indications of residual eccentricity in the waveforms, especially for mass ratios $q=64$ and $q=128$. However, the metadata for these NR simulations does not provide any estimate of initial eccentricities. While it is possible to estimate eccentricities using waveform amplitude or frequencies at the periastron and apostron~\cite{Islam:2021mha,Shaikh:2023ypz}, the shorter duration of the NR data poses significant challenges in obtaining accurate estimates. These methods rely on precise interpolation of the frequencies at the periastron and apostron, which is difficult to achieve in cases where the NR data is limited. For example, we could only find about three apostron and periastron before merger for $q=128$. Consequently, we could not provide any quantitative measurement of the eccentricities.

These limitations should be considered when comparing and validating models in the intermediate mass ratio range using RIT NR data.

%%%%%%%%%%%%%%%%%%%%%%%%%%%%%%%%%%%%%%%%%%%%%%%%%%%%%%%%%%%%%%%%%%%%%%%%%%%%%%%
\bibliography{BHPT_Remnant}
%%%%%%%%%%%%%%%%%%%%%%%%%%%%%%%%%%%%%%%%%%%%%%%%%%%%%%%%%%%%%%%%%%%%%%%%%%%%%%%

\end{document}